\documentstyle[epsf]{mn}
    
\title[On the Viewing Angle Dependence of Blazar Variability]
{On the Viewing Angle Dependence of Blazar Variability}
\author[Avigdor Eldar and Amir Levinson]
{Avigdor Eldar and Amir Levinson\\
School of Physics and Astronomy, Tel Aviv University, 
Tel Aviv 69978, Israel}
\date{Accepted
      Received
      in original form \today}
\pagerange{\pageref{firstpage}--\pageref{lastpage}}
\pubyear{1998}
\begin{document}
\maketitle
\label{firstpage}
\begin{abstract}
Internal shocks propagating through an ambient radiation field, 
are subject to a radiative drag that, under certain conditions,
can significantly affect their dynamics and, consequently, the
evolution of the beaming cone of emission produced behind the 
shocks.  The resultant change of the Doppler factor 
combined with opacity effects leads to a strong dependence of 
the variability pattern produced by such systems, specifically,
the shape of the light curves and the characteristics of correlated
emission, on viewing angle.  One implication is that objects 
oriented at relatively large viewing angles to the observer should
exhibit a higher level of activity at high synchrotron 
frequencies (above the self-absorption frequency) and 
at gamma-ray energies below the threshold energy to pair production,
than at lower (radio/millimeter) frequencies. 

\end{abstract}                                                          
\begin{keywords}
galaxies:jets - radiation mechanisms:nonthermal - relativity
\end{keywords}

\section{Introduction}

The rapid variability often exhibited by blazars and GRBs is widely 
attributed to formation of internal shocks in the expelled
outflow (see e.g., Rees 1978; Romanova \& Lovelace 1997; Levinson 1998, in 
the context of blazars, and Rees \& Meszaros, 1994; Eichler 1994; 
Sari \& Piran 1997, in the context of GRBs).  Internal shocks are produced 
as a result
of unsteadiness of the source, which ultimately leads to 
overtaking collisions of different fluid slabs.  The observed 
variability time associated with a single front is on the order 
of the light travel time across the expelled fluid slab
(provided the cooling time is sufficiently short, and that the slab 
is optically thin and not too thick geometrically), and, therefore,
can be as short as the intrinsic timescale (e.g., the dynamical 
time of the central engine).  If indeed associated with the gravitational
radius of the putative black hole, this timescale comes out to be of order 
milliseconds in the case of GRBs, and minutes to hours in the case 
of blazars, consistent with the temporal substructure seen in these 
two classes of objects (e.g., Wagner, 1997; Ulrich, et al., 1997;
Fishman \& Meegan, 1995).  The fraction of bulk energy that can be
dissipated behind the shocks and, provided the cooling time is sufficiently 
short, radiated away, depends mainly on the difference in Lorentz 
factors of the colliding shells.  In scenarios that invoke magnetically 
dominated outflows, effective dissipation of the shocked magnetic field 
is also required for high radiative efficiency (e.g, Levinson \& Van Putten
1997).

The dynamics of internal fronts and the resulting variability 
pattern depend on the parameters of the expelled fluid and on
environmental conditions.  In particular, in situations whereby
the front moves through an intense, roughly isotropic (in 
the frame of the central engine) radiation field, as in 
ERC models, it will be subject to a radiative drag that can
affect its dynamics and emission considerably.  
In the model considered here (see, Levinson, 1998 [Paper I]; 
Levinson 1999a [Paper II] for details), the created front 
is assumed to be adiabatic initially.  During the adiabatic phase, shortly 
after its creation and prior to 
the onset of radiative losses, the front moves at some constant velocity 
intermediate between that of the colliding fluid slabs, such that the net 
momentum flux incident through the shocks and, consequently, the net 
force exerted on the front by the ``push'' of the exterior fluids vanish. 
A fraction of the energy dissipated behind the shocks is tapped for the 
acceleration of electrons to nonthermal energies and the rest 
to heat the front.  The injected electrons then cool adiabatically, 
owing to the front expansion, and radiatively through synchrotron emission
and inverse Compton scattering of external, soft photos (ERC emission).  
Since the synchrotron emission is isotropic in the comoving frame it 
does not contribute to momentum losses.  However, the ERC emission, which is 
highly beamed in the front frame, gives rise to a radiative drag that 
leads to deceleration of the front during the rise of the emitted
ERC power.  The increasing radiative friction is balanced by an excess
momentum transfer to the front from the fluid behind it (the 
fast fluid).  The reason is that as the front decelerates the 
relative velocity between the front and the fast fluid and, hence, the 
net momentum flux incident through the reverse shock increases, while
the net momentum flux incident through the forward shock decreases.
The decelerating front will reach its minimum Lorentz factor
roughly when the total ERC power radiated by the front peaks, provided 
the shock crossing time of the expelled fluid slabs is sufficiently 
long. If the intensity of external radiation declines with radius, as 
envisaged here, then the ERC flux radiated by the front and the 
associated drag will decline as the front moves outwards.  This 
would then lead to re-acceleration of the front, following peak emission, 
until its initial speed and structure are restored (or until shock 
crossing of the fluid slabs is completed).  The external,
soft photons also contribute a large pair production opacity that 
can lead to absorption of escaping gamma rays and the initiation
of pair cascades inside and ahead of the front at early times.  This
, in turn, will affect the evolution of the high energy spectrum.  The 
synchrotron opacity also changes with time, owing to the front expansion. 
We stress that the variability in this model reflects the dynamics of
the front and the radial profiles of the magnetic field and 
external photon intensity, and is not due to any explicit time 
changes of the front parameters or particle injection.  To be 
concrete, in the case of long outbursts the variability
is caused by a change in the Thomson and synchrotron opacities during 
the course of the front that result from the radial variations of 
external photon intensity and magnetic field.  In the case of short
outbursts the variability is due to shock crossing (after which 
energy deposition in the front ceases) and the subsequent cooling of 
the hot fluid slabs.
The combination of time varying Lorentz factor and optical depth
effects should give rise to a certain dependence of the variability pattern
on the viewing angle (Levinson 1999b).  It is the purpose of this paper 
to explore this orientation effect within the framework of the radiative 
front model developed in Papers I and II.

\section{Intensity and Observed Flux}

In Papers I and II we analyzed the structure and dynamics of 
a radiative front and computed the temporal evolution of the 
angle averaged flux radiated during its course.  In this section we 
generalize our previous treatment to allow for the calculation 
of the angular dependence of the emission.  The structure and
dynamics of the front are computed as before using the model
developed in Paper I, but the equations governing the 
evolution of the radiated flux have been modified in a manner 
described below.  The model assumes
that a constant fraction of the power dissipated behind the 
shocks is injected in the form of nonthermal electron distribution.
The rate of energy dissipation inside the front depends, in turn, 
on the relative velocities 
between the exterior fluids and the front and is computed self consistently.
The electron acceleration time is assumed to be much shorter than the 
cooling and light crossing times, so that it does not affect the
evolution of the radiated ERC and synchrotron spectra.
The energy distributions of electrons, gamma-rays  and synchrotron 
photons inside the front are computed by solving the appropriate 
kinetic equations,
which are coupled to the MHD equations governing the front dynamics
through the injection term and the energy and momentum loss terms. 
As shown in Papers I and II, the energy distribution of emitting 
electrons is determined essentially by the pair cascade and cooling processes 
and is insensitive to the form of injected electron spectrum, provided
the acceleration process is efficient and that the Thomson opacity 
contributed by the external radiation field
is large enough.  In the examples presented below the injected
spectrum was taken to be a power law with roughly equal energy injection
rate per log energy.

We approximate the front (see fig. 1) as a cylindrical section with 
an axial length
$\Delta X$ and cross sectional radius $d$, and denote by $\beta_c$,
$\beta_{s+}$ and $\beta_{s-}$ the velocity of the front,
the forward and reverse shocks with respect to the injection frame,
respectively, and by $\Gamma_c$, $\Gamma_{s\pm}$ the corresponding 
Lorentz factors.  We suppose that in addition to its radial expansion,
which is computed from the model, the front expands also sideways
at some velocity, taken to be a fraction $\psi <<1$ of $\beta_c$.
This then yields $d(t)=\psi r(t)$, where $r(t)=r_o+c\int{\beta_c dt}$
is the position of the front (more precisely, the contact discontinuity)
at time $t$.  We stress that the shock geometry invoked here is
probably unrealistic, since the perimeter of the forward shock moves
at velocity $\beta_c \sqrt{1+\psi^2}$ that may violate causality under 
certain choice of parameters.  The shock is more likely to be curved,
or even corrugated as a result of instabilities.  We do not expect,
however, our results to be strongly dependent upon shock
geometry, but merely on the characteristic velocities.  
We further assume that the electron distribution is 
isotropic and homogeneous inside the front, and
denote by $j_{\nu}(\mu,t)$ and $\kappa_{\nu}(\mu,t)$ the emission 
and absorption coefficients of some radiation process (e.g., synchrotron 
or ERC emission), as measured in the injection frame.  In the case of 
synchrotron emission, the latter 
assumption implies that $j_{\nu}$ and $\kappa_{\nu}$ are also isotropic 
and homogeneous inside
the front (but not necessarily the intensity).  This is not true, however, 
for the ERC emission (Dermer, 1995), since the comoving distribution 
of scattered photons is highly anisotropic.

\begin{figure}
\centerline{\epsfxsize=3.3in\epsfbox{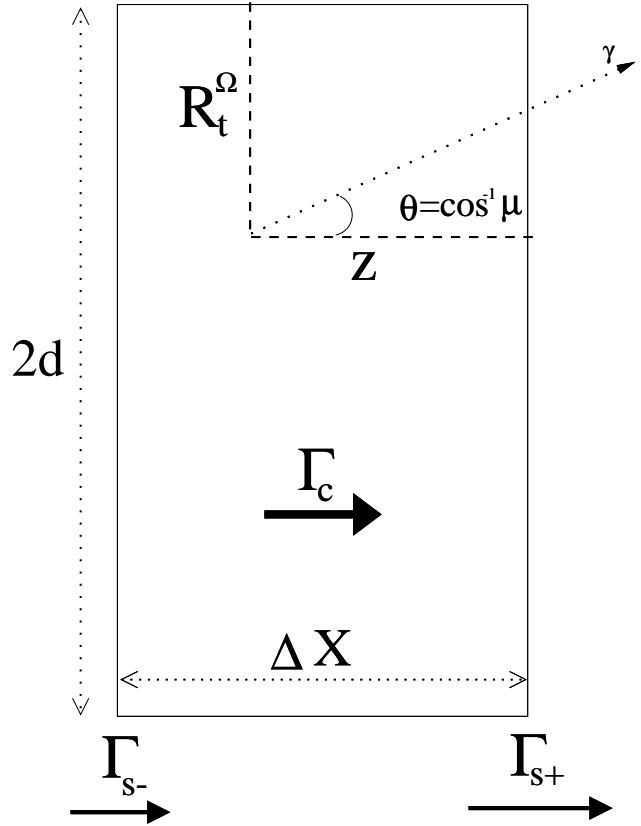}}
\caption{Schematic illustration of the front structure.}
\label{maghistograms}
\end{figure}

Consider now a photon emitted at time t at some position inside the front 
in the direction $\hat{\Omega}$, and denote by $z_t$ the distance 
between the location from which the photon is emitted and the forward shock,
and by $R_t^{\Omega}$ the distance along the projection of $\hat{\Omega}$ 
on the plan perpendicular to the front axis (that is, along the vector 
$\hat{\Omega}-\mu\hat{z}$, where $\mu=\hat{\Omega}\cdot\hat{z}$) between 
the emission point and the front boundary (see fig. 1).
Now, the photon (if not absorbed) will escape from the front through 
the forward shock if c$\bar{t}\sin\theta<R_t^{\Omega}$ 
(ignoring the transverse expansion of the front), and through the sides 
when the opposite inequality holds.  Here $\sin\theta=\sqrt{1-\mu^2}$, 
and $\bar{t}$ is given implicitly by $\bar{t}=(z_t/c+\int^{\bar{t}}
{\beta_{s+} dt})/\mu$.  In terms of the time averaged shock 
velocity, $\bar{\beta}_{s+}$, one obtains, $c\bar{t}=z_t/(\mu-
\bar{\beta}_{s+})$, so that the condition that the photon will escape
through the forward shock reads: $z_t<R_t^{\Omega}(\mu-\bar
{\beta}_{s+})/\sin\theta$.  Note that photons emitted into directions $
\mu<\bar{\beta}_{s+}$ can only escape from the sides.

The intensity crossing the front boundary along the direction 
$\hat{\Omega}$, at some position $\Sigma$ on the boundary 
surface, is given, under the above assumptions, by 
\begin{equation}
I_{\nu}(\Sigma,\hat{\Omega},t)=\int_{0}^{t_{\Sigma}}{j_{\nu}
(\hat{\Omega},t-\chi)e^{-\tau_{\nu}(\chi)}cd\chi},
\label{eq:I}
\end{equation}
where $\tau_{\nu}(\chi)=c\int_{0}^{\chi}{\kappa_{\nu}(\chi')d\chi'}$ is the 
optical depth traversed by a photon emitted at time $t-\chi$,
and $t_{\Sigma}$ is the light crossing time of the expanding front 
along the corresponding ray, measured in the frame of the central 
engine.  The flux emitted from the boundary 
surface at some location is given by $I_{\nu}(\Sigma,\hat{\Omega},t)
(\hat{\Omega}\cdot\hat{n}-\hat{\beta}_{\Sigma}\cdot\hat{n})$, where 
$\hat{n}$ is the normal to the surface, and ${\beta}_{\Sigma}$
is the velocity of the surface at the corresponding position (for  
instance ${\beta}_{\Sigma}=\beta_{s+}\hat{z}$ for any point on 
the forward shock surface).  This must be multiplied by the time 
dilation factor, $(1-\beta_{\Sigma}\cdot\hat{\Omega})^{-1}$, in order to 
obtain the observed flux.  One finds,
\begin{equation}
{\cal F_{\nu}}=\frac{(1+Z)}{D_L^2}\int_{\Sigma}{I_{\nu}(\Sigma,
\hat{\Omega},t)\frac{(\hat{\Omega}\cdot\hat{n}-\hat{\beta}_{\Sigma}
\cdot\hat{n})}{1-\beta_{\Sigma}\cdot\hat{\Omega}}d\Sigma}.
\label{eq:F}
\end{equation}
Here $D_L$ is the luminosity distance and $Z$ is the corresponding 
redshift.  We note that the time evolution of $\beta_{\Sigma}$
and the remaining front parameters is computed using the front equations 
derived in Paper I.  It is also worth-noting that the observed time change,
$t_{obs}=\int(1-\beta_{\Sigma}\cdot\hat{\Omega})dt$, may be 
different for different emitting surfaces, so that the observed time 
structure may reflect, to some extent, the geometry of the front.  
The above treatment modifies the analysis presented in Papers I and II to 
account for retardation effects associated with the front expansion and 
with rapid time changes of the emissivity and opacity.  We find that this
modification does not give rise to significant alterations of 
the results obtained in Papers I and II for the evolution of the angle  
averaged flux, but does improve slightly the calculations of the light 
curves observed at relatively large viewing angles.  

As a simple example consider a non-expanding blob moving with a
velocity $\beta_c$, and let the volume
emissivity be time independent.  Then $\beta_{\Sigma}=\beta_c\hat{z}$,
and equations (\ref{eq:I}) and (\ref{eq:F}) yield for the optically 
thin flux, ${\cal F_{\nu}}\propto
\frac{j V}{(1-\beta_c\mu)}$, where $V=ct_{\Sigma}\int d\Sigma$ is the 
volume of the front.  In terms of the comoving volume, $V'=\Gamma_c V$,
and comoving emissivity,
the observed flux reduces to the familiar expression:    
${\cal F_{\nu}}=\frac{(1+Z)}{D_L^2}{\cal D}^3_{c}j'_{\nu} V'$, with
${\cal D}_{c}=[\Gamma_{c}(1-\beta_{c}\mu)]^{-1}$ being the 
corresponding Doppler factor.  As a second example, consider the flux
emitted from an expanding, stationary front in the forward direction, 
viz., $\mu=1$.  In that case the entire flux is emitted through the 
forward shock.  The corresponding time dilation factor is then
$(1-\beta_{s+})^{-1}$, and the light crossing time along the front's
axis is $t_{\Sigma}=\bar{t}=\Delta X/(1-\beta_{s+})$.  
From equations (\ref{eq:I}) and (\ref{eq:F}) we obtain, again in
the optically thin limit, $${\cal F_{\nu}}=\frac{(1+Z)}{D_L^2}{\cal 
D}_{s+}{\cal D}_{c}^2(\Gamma_{s+}/\Gamma_c) V' j_{\nu}',$$
with ${\cal D}_{s+}$ being the Doppler factor associated with the 
forward shock. 
This example illustrates the effect of the expansion on the emitted 
flux.  The enhancement of the forward flux by a factor 
$(\Gamma_{s+}/\Gamma_c)^2$ is entirely due the growth of the front's 
volume. 

The integration of eq. (\ref{eq:I}) for a specific process requires 
the determination of the corresponding volume emissivity $j_{\nu}$.
The determination of the synchrotron emissivity is straightforward,
since the emission is isotropic in the front frame,
and is described in Paper II.  The ERC emissivity is calculated using 
the head-on approximation; that is, the direction of scattered photon 
is taken to be along the direction of the scattering electron.  The 
electron distribution in the injection frame, denoted by $n_e(E_e,\mu,t)$, 
is obtained at each time step by appropriate Lorentz transformation of
the comoving electron distribution which, as mentioned above,
assumed to be isotropic .  Using this approximations the ERC emissivity 
can be expressed as,
\begin{equation}
j_{ERC}(E_{\gamma},\mu,t)=\int{n_e(E_e,\mu,t)\eta_{c\gamma}
(E_{\gamma},E_e,t)dE_e}
\end{equation}
where $\eta_{c\gamma}$ is the corresponding redistribution function,
and is given explicitly in Blandford \& Levinson (1995).  Note that 
$\eta$ is independent of $\mu$ by virtue of the assumed isotropy of 
the ambient radiation field.  Finally, since we consider only cases 
for which gamma-ray production is dominated by ERC emission, we 
neglect the SSC emissivity in eq. (\ref{eq:I}) (see \S 4 for further
discussion).  
                
\section{Results}

Equations (\ref{eq:I}) and (\ref{eq:F}) and the front equations 
derived in Paper I were integrated for different values of 
$\mu$. 
In the following examples, the Lorentz factors of the fluids ahead 
and behind the front and
the rest-frame Alfven 4-velocity have been chosen to be, respectively,
5, 20, and 10.  The magnetic pressure has been taken to decline as
$r^{-p}$, and the intensity of background radiation as $f(r)/r^{2}$.
A rapid magnetic field dissipation inside the front with the same decay 
constant as in the previous papers has been invoked.  As a check, we
computed the angular distribution of observed flux for a front with
roughly time independent Lorentz factor (low radiative efficiency case),
and compared the results with 
the analytic expressions, ${\cal F}\propto {\cal D}^{\delta}$, with 
$\delta=3+\alpha$ and $\delta=4+2\alpha$ for synchrotron and ERC 
emission, respectively (Dermer 1995).  The analytic results were 
reproduced to a very good accuracy by the numerical model.   

Fig. 2 depicts the time profile of the front Lorentz factor, 
$\Gamma_c$, obtained for sufficiently long outbursts, in the sense
that the shock travel time across the fluid slab is longer than the
time change associated with the radial variations of magnetic field
and ambient radiation intensity (see Paper I for more details).  The 
radial profiles of magnetic field and external radiation field 
in this example are $r^{-2}$ ($p=2$, $f(r)=1$).  A steeper profile
of the ambient radiation intensity leads quite generally to a larger
acceleration of the front following peak emission, owing to the faster
decrement of the radiative friction experienced by the front. 
The times of peak emission at different 
bands, and the corresponding Lorentz factors are indicated in the
figure, and it is seen that the emission at different energies 
has different Doppler factors.  This is essentially a consequence
of the combination of dynamical and opacity effects.  In addition to
this dependence of Doppler factor on wavelength, there will also be a 
difference in the beaming patterns of synchrotron
and ERC emission, owing to the difference in angular dependence of
the corresponding volume emissivities (see below).

\begin{figure}
\centerline{\epsfxsize=3.3in\epsfbox{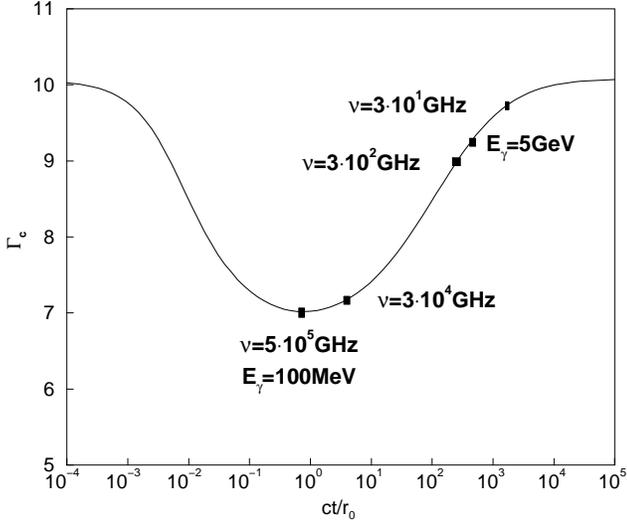}}
\caption{Time profile of the front Lorentz factor obtained for
sufficiently long outbursts (see text). The times of peak emission
at different energies are indicated}
\label{2}
\end{figure}

The resultant beaming patterns at different energies are delineated in fig. 3,
where the relative dependences of the peak fluxes (normalized to their values
at $\mu=1$) on viewing angle are exhibited.  The two gamma-ray bands 
shown correspond to the total gamma-ray flux above and below the 
threshold energy at which the pair production opacity (at $r=r_o$) 
is roughly unity.  The difference in beaming 
patterns between synchrotron and ERC emission is clearly seen.  Also evident
is the stronger dependence of the optically thick emission on viewing angle.
Although a model for the quiescent emission is required in order to 
calculate the amplitude of variations, the dependence of observed flux 
on $\mu$ shown in
fig. 3 suggests that objects oriented to the observer at relatively 
large viewing angles may have preferentially larger 
amplitude outbursts at short wavelengths (optical/UV), and for small enough
viewing angles also at gamma-ray energies below the break of the gamma-ray
spectrum. 
It is also conceivable that intense outbursts at these energies
will be followed by no activity at all at low frequencies.

\begin{figure}
\centerline{\epsfxsize=3.3in\epsfbox{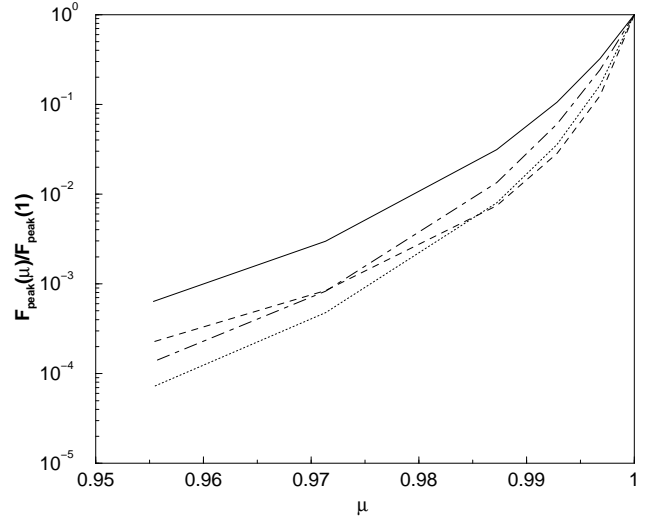}}
\caption{Peak fluxes (normalized to their values at 
$\mu=1$) versus cosine of the viewing angle, at different energy 
bands: $5\times10^5$ GHz (with a logarithmic energy interval) 
(solid line), $3\times10^2$ GHz (dashed line),
total gamma-ray flux in the energy interval 10 MeV to 1 GeV 
(dotted-dashed line), and 1 to 10 GeV (dotted line).  An ambient 
radiation intensity profile with
$f(r)=1$ (corresponding to a $r^{-2}$ profile) was used in this 
calculation}
\label{3}
\end{figure}

The angular dependence of observed flux reflects, in this model, 
also the radial profiles of magnetic field and ambient radiation 
intensity.  This is because both, the opacity and the front 
dynamics depend on the radial decrement of these quantities.  
Quite generally we find that the dependence of the ratio of optically thin 
and optically thick synchrotron fluxes on viewing angle becomes 
stronger for steeper ambient intensity profiles.
The results obtained for ambient radiation intensity with an 
exponential profile, $f(r)=\exp(-r/r_1)$ (with $r_1=10r_o$ in this 
example), that may reflect the density profile of the surrounding gas 
that re-processes or scatters the nuclear radiation across the front, 
are shown in Fig. 4.  As seen, the dependence of the optically thin
fluxes on viewing angle is insensitive to the form of $f(r)$.  However, the 
peak of the optically thick synchrotron flux declines more steeply 
with increasing
viewing angle.  This is because the front re-accelerates much faster 
in this case, owing to the rapid drop in radiative drag, so that the
Doppler factor of the optically thick bands changes more rapidly. 
The behavior of optically thick ERC emission is more complicated; 
steeper ambient intensity profiles render the 
pair production opacity ahead of the front smaller which, in turn, leads to 
earlier ERC emission.  This counteracts the effect associated with the
faster acceleration and, therefore, the angular 
dependence of this component is more sensitive to the choice 
of parameters.  In fact, in the example shown in Fig. 4 the decline of 
the total gamma-ray
flux above the threshold energy to pair production at $r_o$
is slower in the case of ambient intensity 
with an exponential profile, in contrast to the behavior of the low-frequency 
synchrotron emission.  

The radial profile of the magnetic field affect mainly the synchrotron
flux.  Steeper profiles give rise to shorter delays of the low-frequency 
emission and, depending on the front acceleration scale, lead to 
a weaker dependence of the peak flux on viewing angle.

\begin{figure}
\centerline{\epsfxsize=3.3in\epsfbox{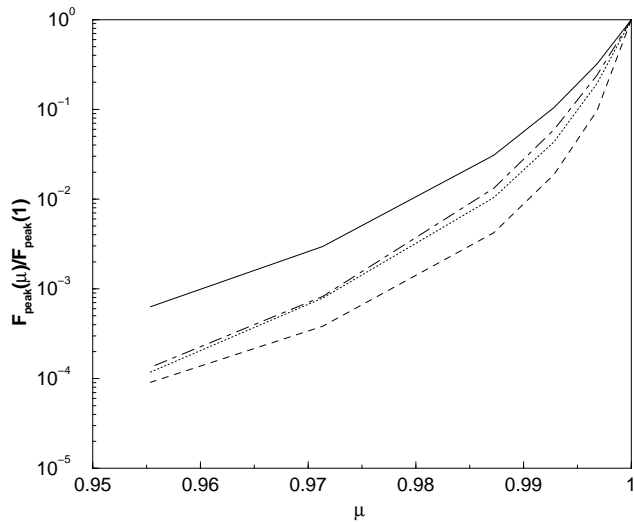}}
\caption{Same as fig. 3 but for $f(r)=exp(-r/r_1)$, with $r_1=10r_o$}
\label{4}
\end{figure}

Fig. 5 presents sample light curves at various energies, computed 
for different viewing angles.  As seen, for optically thick bands the shape
of the light curve depends on $\mu$, tending towards a much steeper decline
at larger values of $\mu$, thereby leading to a light curve that appears more 
symmetric.  Moreover, the time of peak emission and flare duration decrease
with increasing viewing angle, as can be seen from the figure.  This is 
again due to the evolution of the 
Doppler factor caused by the re-acceleration of the front.  The optically
thin bands show little dependence of flux decay time on viewing angle, as
expected.  We note that in situations where the ejected slabs are thin enough, 
or the ambient intensity has a much steeper radial profile,
the flares will also tend to have a roughly symmetric shape.

\begin{figure}
\centerline{\epsfxsize=3.3in\epsfbox{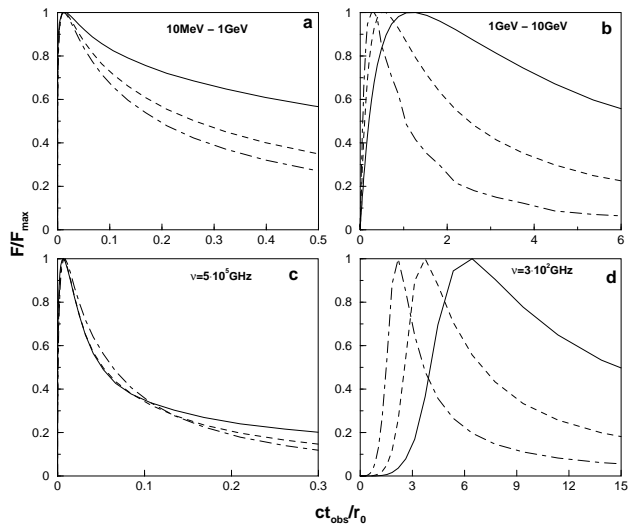}}
\caption{Sample light curves produced by the model, for $\mu=1$ (solid
line), $\mu=0.99$ (dashed line), and $\mu=0.97$ (dotted-dashed line). 
Each window corresponds to a given energy band (as indicated)}
\label{5}
\end{figure}

In the case of sufficiently short outbursts, shock crossing is completed
before front re-acceleration sets in.  The dependence of the variability 
pattern on viewing angle would then be different than for long outbursts,
and may depend on the cooling time of emitting electrons in the heated 
fluid slab.
If only the fast slab is short, the reverse shock will decay quickly while
the forward shock will continue to propagate outward, similar to a 
GRB blast wave.  This may give rise to somewhat different characteristics 
of the low-frequency emission.   If both slabs are thin, then a lack of
activity at long wavelengths, and a lower energy cutoff at gamma-ray
energies are expected (see Paper II for a detailed discussion).  
The temporal behavior will be further complicated in situations in
which multiple fronts with small enough duty cycle are created.  This can lead
to blending of different flares (as often seen at radio wavelengths; e.g., 
Aller, et al. 1985; Valtaoja, et al. 1999) and, in the case of 
ejection of a thin 
slab, to a collision of the decelerating slab, following shock crossing, 
with a newly expelled one, and the ultimate formation of a new shock in
the radiating slab.  Such episodes are far more difficult to simulate.
Nonetheless, rapid, large amplitude flares should
reflect the features associated with a single front.
                
\section{Conclusions}

We have considered the angular dependence of the observed variability
pattern produced by internal fronts propagating through an ambient
radiation field.  We have shown that, for sufficiently long outbursts,
the combination of dynamical effects caused by the radiative drag 
and optical depth effects, gives rise to a strong dependence of the 
observed flare's properties on source orientation.

To be more concrete, the shape of the light curves of optically thick emission
reflects, at sufficiently large viewing angles, the temporal evolution of 
the Doppler factor, and at small viewing angles the evolution 
of the radiated power density.  As a consequence, the time of peak emission
and flare duration decrease with increasing viewing angle.  Moreover,
the flare appears to be more symmetric at larger viewing angles.  The
time evolution of optically thin emission is insensitive to source 
orientation, since it originates when the front is near its minimum
Lorentz factor.

The time evolution of the beaming factors also renders the 
characteristics of correlated emission sensitive to source orientation.
Because the source is inhomogeneous, the fluxes at different 
energies originate from different locations 
along the course of the front and, therefore, have different beaming 
cones due to the varying Lorentz factor.  As a result the ratio of 
peak fluxes of the low-frequency (self-absorbed) and high-frequency
synchrotron emission decreases with increasing viewing angle. 
The beaming cone of ERC emission is narrower than that of the 
synchrotron emission, leading to a somewhat more sensitive dependence
of the gamma-ray flux on viewing angle than the high-frequency synchrotron
flux.  One implication of the above results is that sources that are 
oriented at relatively large viewing angles to the observer may exhibit 
events whereby strong gamma-ray and/or IR-optical-UV flares are 
followed by little or no activity of the low frequency 
(radio-to-millimeter) flux.

Finally, we note that the neglect of SSC emission may not be justified 
at large viewing angles, even in cases where the total power radiated
is dominated by ERC emission.  This is because the beaming cone of
ERC photons is narrower than that of SSC photons (Dermer 1995).  Thus, 
it is conceivable that in some regime of parameter space, the origin
of observed gamma-rays also depends on the orientation of the source 
with respect to the observer.

This research was supported by The Israeli Science Foundation. 

\break
 
\label{lastpage}
                  
\end{document}